\documentclass[12pt]{article}
\usepackage{graphicx}
\usepackage{a4wide}
\usepackage{amssymb}
\begin{document}
{\renewcommand{\thefootnote}{\fnsymbol{footnote}}
\hfill  IGC--08/1--3\\
\medskip
\begin{center}
{\LARGE  Quantum nature of cosmological bounces }\\
\vspace{1.5em}
Martin Bojowald\footnote{e-mail address: {\tt bojowald@gravity.psu.edu}}
\\
\vspace{0.5em}
Institute for Gravitation and the Cosmos,\\
The Pennsylvania State
University,\\
104 Davey Lab, University Park, PA 16802, USA\\
\vspace{1.5em}
\end{center}
}

\setcounter{footnote}{0}

\newtheorem{theo}{Theorem}
\newtheorem{lemma}{Lemma}
\newtheorem{defi}{Definition}

\newcommand{\proofend}{\raisebox{1.3mm}{\fbox{\begin{minipage}[b][0cm][b]{0cm}
\end{minipage}}}}
\newenvironment{proof}{\noindent{\it Proof:} }{\mbox{}\hfill \proofend\\\mbox{}}
\newenvironment{ex}{\noindent{\it Example:} }{\medskip}
\newenvironment{rem}{\noindent{\it Remark:} }{\medskip}

\newcommand{\case}[2]{{\textstyle \frac{#1}{#2}}}
\newcommand{\lP}{\ell_{\mathrm P}}

\newcommand{\md}{{\mathrm{d}}}
\newcommand{\tr}{\mathop{\mathrm{tr}}}
\newcommand{\sgn}{\mathop{\mathrm{sgn}}}

\newcommand*{\R}{{\mathbb R}}
\newcommand*{\N}{{\mathbb N}}
\newcommand*{\Z}{{\mathbb Z}}
\newcommand*{\Q}{{\mathbb Q}}
\newcommand*{\C}{{\mathbb C}}

\begin{abstract}
  Several examples are known where quantum gravity effects resolve the
  classical big bang singularity by a bounce. The most detailed
  analysis has probably occurred for loop quantum cosmology of
  isotropic models sourced by a free, massless scalar. Once a bounce
  has been realized under fairly general conditions, the central
  questions are how strongly quantum it behaves, what influence
  quantum effects can have on its appearance, and what quantum
  space-time beyond the bounce may look like. This, then, has to be
  taken into account for effective equations which describe the
  evolution properly and can be used for further phenomenological
  investigations. Here, we provide the first analysis with interacting
  matter with new effective equations valid for weak self-interactions
  or small masses.  They differ from the free scalar equations by
  crucial terms and have an important influence on the bounce and the
  space-time around it.  Especially the role of squeezed states, which
  have often been overlooked in this context, is highlighted. The
  presence of a bounce is proven for uncorrelated states, but as
  squeezing is a dynamical property and may change in time, further
  work is required for a general conclusion.
\end{abstract}

\section{Introduction}

Classical cosmology as a description of the early universe is plagued
by the big bang singularity. In this extreme high energy and high
curvature regime quantum space-time is expected to take over the
reigns from general relativity in a way which will hopefully cure the
singularity problem. A common expectation is that the universe,
instead of collapsing completely, will bounce at a minimum non-zero
volume and thus connect to a well-defined pre-big bang state. Mediated
by quantum gravity, such a picture would exploit the changes in forces
and the dynamics which quantum physics should imply. If this can be
realized, a central question remains: Do the quantum forces that have
been summoned to fight the classical singularity leave space-time
otherwise intact, or will space-time remain in a fluctuating turmoil
long after the immediate danger of divergences has passed?  Generally,
strong quantum effects as they may be necessary to prevent the
singularity lead to large fluctuations. Then, a genuine quantum
space-time has to be dealt with, which may not be near a smooth
bouncing effective picture.

Detailed examples have been provided by loop quantum cosmology
\cite{LivRev}, where homogeneous models are governed by a non-singular
difference equation for quantum states \cite{Sing,IsoCosmo,Bohr,BSCG}. The
general difference equation, however, does not provide direct insights
into intuitive geometric properties around the classical singularity;
in general the transition may not be semiclassical at all. But there
is a class of models which are more easily accessible: isotropic
models sourced solely by a free, massless scalar. Initial numerical
calculations \cite{QuantumBigBang,APS} have suggested that the
singularity is avoided by a surprisingly smooth bounce of a wave
packet which remains sharply peaked and fluctuated before the bounce
very nearly as it does afterwards. ``Before'' and ``after'' here
refers to a global internal time which in these models is provided by
the value of the scalar.

Upon closer examination, however, two new key features materialized:
\begin{itemize}
\item In a suitable factor ordering of the Hamiltonian constraint
  operator, a flat isotropic model with a free, massless scalar is a
  solvable system in which quantum back-reaction does not occur
  \cite{BouncePert}. This is the case even in the bounce phase of the
  loop quantization. Thus, quantum variables such as fluctuations or
  higher moments of a state do, although they are themselves dynamical
  and change in time, not influence the evolution of expectation
  values. It is this feature which makes the models highly controlled,
  but also very special. In quantum mechanics, for instance, this
  behavior is realized only for the harmonic oscillator. When
  interactions, or deviations from isotropy, are included quantum
  back-reaction does result and the evolution will become more quantum
  in the sense that it depends on how the quantum variables behave.
  Quantum correction terms due to coupling between expectation values
  and fluctuations would at first not be large for evolution starting
  with a semiclassical state, as it should be realized at large volume
  of a universe. However, states spread and deform from an initial
  Gaussian distribution. Accordingly, quantum variables generically
  grow in time and with them quantum corrections. This may take time,
  but can be significant in long-term evolution. Cosmology is a study
  of the system with the longest possible evolution times, and
  especially regarding the big bang singularity the correct question
  to ask is: What is the behavior of a universe at small volume,
  described by a {\em generic} quantum state as it has evolved from a
  semiclassical state at large volume? This question cannot be
  answered if the behavior of the quantum system is known only for
  semiclassical states and rather short evolution times. For all we
  know, the universe near the big bang can be in a highly quantum
  state. Thus, quantum back-reaction is a serious player in this
  regime, and one has to understand how it may affect the bounce of
  the free solvable model.
\item Even within the solvable model, fluctuations before and after
  the bounce can be very different from each other
  \cite{BeforeBB,Harmonic}.\footnote{Some of the recent literature on
  this topic has witnessed a certain amount of reluctance to accept
  this property. For instance, it has been claimed in
  \cite{BounceRecall} that fluctuations of a state which is
  semiclassical at large volume have to be very nearly symmetric
  around the bounce.  What was actually derived there was an upper
  bound for the difference before and after the bounce of fluctuations
  of volume relative to the total volume. However, the discussion
  ignored the size of these relative fluctuations and overlooked the
  fact that each of the terms in the difference, not just the
  difference, is smaller than the upper bound provided. Thus, the
  inequality does not at all restrict the asymmetry. One can easily
  see that the numerical example provided in \cite{BounceRecall}
  allows for ratios of relative fluctuations before and after the
  bounce as large as $10^{28}$. Judging from this estimate alone, a
  state before the big bang may be highly non-semiclassical even if it
  becomes very semiclassical after the big bang. The detailed bounds
  provided in \cite{Harmonic} are much sharper, but still leave room
  for asymmetric fluctuations.}  This asymmetry is not a consequence
  of quantum back-reaction but of squeezing. As the precise state is
  unknown, there is no strong control on how it may be squeezed even
  when it is semiclassical. Accordingly, no strong control on
  fluctuations during and before the big bang exists. Still, for any
  such state there is a bounce in this free model provided that the
  state is semiclassical at one time.
\end{itemize}

The formulation in \cite{BouncePert} as a solvable model allows the
derivation of precise effective equations in this case, which easily
shed light on the evolution of a quantum space-time and its
properties. If the model is perturbed by including a small matter
potential, perturbative effective equations can still be derived in an
expansion \cite{BouncePot}. In this paper, we use these equations to
show that the bounce is in fact affected by interactions and determine
how this is related to the possible squeezing of the state. Our
analysis is thus based on a combination of the two key results which
are realized in isotropic, free scalar models. We provide a new
effective equation which, unlike previous examples, is valid for
a massive or self-interacting scalar. It contains additional
correction terms which are especially important near a would-be bounce.

\section{Loop quantum cosmology}

In loop quantum cosmology, the basic pair of canonical variables is
formed by the square $p=a^2$ of the scale factor $a$ and extrinsic
curvature, i.e.\ the time derivative $c=\gamma\dot{a}$. Their Poisson
bracket is $\{c,p\}=8\pi\gamma G/3$. These are isotropic reductions of
a densitized triad $E^a_i=p\delta^a_i$ and the Ashtekar connection
$A_a^i=c\delta_a^i$ and enter the basic operators of a loop
quantization. Although $p$ can take both signs, signifying the two
possible orientations of the triad, we will mainly restrict attention
to positive values.

While $p$, just as the densitized triad of the full theory, can
readily be quantized through flux operators, there is no operator for
the connection components. Instead, holonomies of the connection are
represented in a well-defined way, which are obtained after
integrating the connection along curves and exponentiating the result.
In an isotropic model, it is not obvious what the length of the curve
used in the integration should be, thus giving rise to quantization
ambiguities which can only be resolved in a detailed reduction of the
isotropic model from the full quantum theory. (See
\cite{SymmRed,AnisoPert,SymmQFT,Reduction,InhomLattice,SymmStatesInt,Rieffel}
for work in this direction, which remains incomplete.) What
length\footnote{The notion of length here only refers to coordinates
and so it may seem that it is not covariant. However, no such problems
or even issues with potential gauge artifacts arise as one can see
upon closer examination \cite{InhomLattice}. To see this, we have to
refine our definition of the basic variables as $p=V_0^{2/3}a^2$ and
$c=V_0^{1/3}\gamma\dot{a}$ where $V_0$ is the coordinate volume of a
region chosen for the spatial integration of the symplectic structure
and the Hamiltonian \cite{Bohr}. Holonomies then take the form
$\exp(i\ell_0V_0^{-1/3}c)$ where $\ell_0$ is the coordinate length of
an edge as referred to above. This immediately shows that the
expression is not coordinate dependent. It also does not depend on the
choice of $V_0$ since this dependence cancels between $c$ and
$V_0^{-1/3}$. Instead of referring to a coordinate quantity, the
length factor can be interpreted as ${\cal N}^{-1/3}$ where ${\cal
N}=V_0/\ell_0^3$ is roughly the number of vertices of an underlying
spin network state in the volume $V_0$. This immediately shows the
dynamical aspect of a pre-factor $f(p)$ in $\exp(if(p)c)$, which
includes the factor $V_0$, if the graph is being refined during
evolution. In this way, ${\cal N}(p)$ will be the number of vertices
of the state at volume $p^{3/2}$. In this context, notice that a
solution to the Hamiltonian constraint equation in general is a
superposition of many states associated with different graphs. The
function ${\cal N}(V)$ refers to this relationally as the number of
vertices of graphs in a decomposition as eigenstates of the volume
operator. This indicates how $f(p)$ refers to internal time evolution
in a constrained theory.}  is used should depend on the underlying
state and its scale of discreteness, which in general should be time
dependent. Thus, also the length factor $\mu$ of a holonomy $\exp(i\mu
c)$ in an isotropic model could be a function of time, or rather of
the spatial geometry since this determines the evolving scales.

A quantum theory with a time dependent number of degrees of freedom is
difficult to formulate and analyze
\cite{Weiss,UnruhTime,River,EvolvingHilbert,CosConst}, but key
properties can be included in the description of an isotropic model.
Instead of using a fixed $\mu\in{\mathbb R}$, we formulate the
dynamics with holonomies of the form $\exp(if(p)c)$ where $f(p)$ takes
into account a possible dependence of the discreteness scale on the
total size. When quantized, the exponential of the momentum $c$
becomes a shift operator in $p$, rather than a derivative operator.
For $f(p)={\rm const}$, the basic triad variable $p$ would be
equidistantly spaced when acting with powers of holonomies, but not
for a $p$-dependence of $f(p)$.  Of special interest are functions
$f(p)=f_0 p^x$ of power-law form, which can arise from loop quantum
gravity for $-1/2<x<0$ \cite{InhomLattice}.  Several independent
phenomenological and stability arguments prefer a value near the lower
bound $x\approx -1/2$
\cite{APSII,SchwarzN,RefinementMatter,RefinementInflation,Vector,Tensor}.
For arbitrary $x$, a canonical pair is formed by the new variables
$(f(p)c,V)$ with
\begin{equation} \label{V}
 V:=\frac{3p}{8\pi \gamma G(1-x)f(p)}=\frac{3a^{2(1-x)}}{8\pi\gamma G(1-x)f_0}
\quad\mbox{ such that }\quad  \{f_0p^xc,V\}=1\,.
\end{equation} 
(We have $V$ proportional to the spatial volume for the power-law case
$f(p)\propto p^{-1/2}$. In this case, the number of vertices in a
graph would grow in a way proportional to spatial volume: ${\cal
N}(a)= f(a^2)^{-3}\propto a^3$. ) This shows that for a given $x$,
$p/f(p)$ is equidistantly spaced upon action of $\exp(if(p)c)$.

In loop quantum cosmology, we have to quantize the classical Friedmann
equation
\begin{equation} \label{IsoConstr}
 \left(\frac{\dot{a}}{a}\right)^2= 
 \frac{8\pi G}{3}\frac{H_{\mathrm{matter}}(a)}{a^3}= \frac{8\pi G}{3}\rho
\end{equation}
where $H_{\rm matter}$ is the Hamiltonian of matter and $\rho$ its
energy density. This is done via the Hamiltonian constraint, which is
obtained by multiplying the Friedmann equation with $a^3$. For a free,
massless scalar, we have the matter Hamiltonian $H_{\rm
matter}=\frac{1}{2}a^{-3}p_{\phi}^2$ with the scalar momentum
$p_{\phi}$. The gravitational part requires the use of $c$, which in
loop quantum cosmology can only be formulated by means of holonomies
which are periodic functions of their arguments. There is no exact
correspondence, but all we need is a well-defined operator which has
(\ref{IsoConstr}) as its low-curvature limit.  Many examples for such
operators exist, the simplest choice being of the form
\begin{equation} \label{QuantFriedmann}
\frac{3}{8\pi\gamma^2 G}  \widehat{(f(p)^{-2}\sin^2(f(p)c)\sqrt{p})} 
\psi(p,\phi)=
  -\frac{1}{2}\hbar^2 \widehat{p^{-3/2}}
  \frac{\partial^2}{\partial\phi^2} \psi(p,\phi)\,.
\end{equation}
(See e.g.\ \cite{AmbigConstr} for a discussion of some ambiguities in
this case.) We have not yet decided on the factor ordering between $p$
and $c$ on the left hand side, and also the right hand side is not
fully defined as written because $p^{-3/2}$ requires an inverse, which
does not exist for the operator $\hat{p}$ with its discrete spectrum
containing zero in loop quantum cosmology. Its quantization thus has
to be more indirect, which can be done \cite{InvScale} following
general constructions of the full theory \cite{QSDV}. (Also here,
ambiguities arise \cite{Ambig,ICGC} without changing the qualitative
picture.)

\subsection{Difference equation}

Basic operators of the quantization are the flux operator
$\hat{p}|\mu\rangle=\frac{4}{3}\pi \gamma\lP^2\mu|\mu\rangle$ with the
Planck length $\lP=\sqrt{G\hbar}$, whose orthonormal eigenstates
$|\mu\rangle$ with $\mu\in{\mathbb R}$ we are using, and the holonomy
operator $\exp(i\mu' c/2)|\mu\rangle = |\mu+\mu'\rangle$. By
expressing the Friedmann equation in terms of holonomies, which then
become finite shift operators, one thus obtains a difference equation
for wave functions \cite{cosmoIV,IsoCosmo}.

There are several ingredients for an explicit expression of the
difference equation: First, using holonomies for a given refinement
scheme we refer to shift operators $\exp(if(p)c/2)|\mu\rangle \sim
|\mu+F(\mu)\rangle$ of a $\mu$-dependent step-size, where
$F(\mu):=f(\frac{4}{3}\pi\gamma\lP^2\mu)$. (This
action is not precisely written yet due to factor ordering which is
discussed below.) The factor $\sqrt{p}$ in (\ref{QuantFriedmann})
could be quantized directly via $\hat{p}$, but is usually written in a
way closer to what one has in the full theory. The reason is that in
an inhomogeneous expression one would instead have a factor
$E^a_iE^b_j/\sqrt{|\det (E^c_k)|}$ which, due to the inverse of the
determinant, cannot be quantized directly.  Following the same steps
in an isotropic model, one is led to a quantization of $\sqrt{p}$ of
the form $if(p)^{-1} \exp(if(p)c) [\exp(-if(p)c),|\hat{p}|^{3/2}]$
which also happens to be diagonalized by the triad eigenstates
$|\mu\rangle$. (Note, however, that the diagonalization property does
not hold in the full theory \cite{BoundFull,DegFull}.) Finally, the
factors $f(p)^{-2}$ and $p^{-3/2}$ have to be turned into operators.

An explicit realization faces two main issues: (i) the non-equidistancy
of step-sizes and (ii) the factor ordering. The ordering between
$\sqrt{q}$ and $f(p)^{-2}\sin^2(f(p)c)$ can be mimicked from the full
theory \cite{QSDI,QSDII}, which leads one to order the quantization of
$\sqrt{p}$ to the right to obtain the basic Hamiltonian constraint
operator $\hat{H}$ which may in a second step be ordered symmetrically
as $\frac{1}{2}(\hat{H}+\hat{H}^{\dagger})$. However, the factors of
$f(p)$ model the refinement behavior of an underlying lattice state,
and thus do not occur in the full theory where the lattice refinement
is automatically realized by elementary holonomies. Thus, the ordering
in isotropic models is not strictly defined, and one cannot always use
guidance from the full theory. In applications, one should therefore
use only properties which are insensitive to the factor ordering. This
is also useful for the first issue: the non-equidistancy of the
resulting difference equations. As shown in \cite{SchwarzN}, one can
always transform a difference equation of isotropic models to
equidistant form, up to changing the factor ordering. Thus, one can
use preferred orderings in which an equidistant difference equation
results which is much easier to analyze. (See \cite{RefinedNumeric}
for a numerical procedure to analyze non-equidistant difference
equations directly.)

Specifically, we obtain a Hamiltonian constraint operator
\begin{equation}
 (\hat{H}-\hat{H}_{\rm matter})|\mu\rangle = 
\frac{1}{32\sqrt{6}\pi\gamma^{3/2} G}
 (|\mu+F(\mu)|^{3/2}- |\mu-F(\mu)|^{3/2}) F(\mu)^{-3}
 (|\mu+4F(\mu)\rangle- 2|\mu\rangle+ |\mu-4F(\mu)\rangle)
\end{equation}
by following the steps detailed in \cite{IsoCosmo,LivRev}.  A physical
state $|\psi\rangle=\sum_{\mu}\psi_{\mu}(\phi)|\mu\rangle$ has to
satisfy $\hat{H}|\psi\rangle=0$, which for its coefficients translates
into the difference equation
\begin{eqnarray}
&& |\Delta_F(\mu+4F(\mu))|F(\mu+4F(\mu))^{-3}\psi_{\mu+4F(\mu)}(\phi)-
 2|\Delta_F(\mu)| F(\mu)^{-3} \psi_{\mu}(\phi)\nonumber\\
&&+
 |\Delta_F(\mu-4F(\mu))|F(\mu-4F(\mu))^{-3}\psi_{\mu-4F(\mu)}(\phi)=
 -64\sqrt{6}\pi\gamma^{3/2} G \hat{H}_{\rm matter}(\mu) 
\psi_{\mu}(\phi) \label{DiffEq}
\end{eqnarray}
where $\Delta_F(\mu):=|\mu+F(\mu)|^{3/2}- |\mu-F(\mu)|^{3/2}$. (The
Wheeler--DeWitt equation in a particular factor ordering is reproduced
in the continuum or large volume limit where $\mu\gg F(\mu)$ at large
$\mu$ \cite{SemiClass}.)

While such a difference equation of varying step-size $F(\mu)$ is
difficult to solve and analyze, it may be transformed to an
equidistant equation up to factor ordering. For this, we use
$\psi_{\mu+F(\mu)}= \tilde{\psi}_{\tilde{\mu}+1}- \frac{1}{2}F'(\mu)
\tilde{\psi}'+ \cdots$ where omitted terms are higher derivatives of
$\tilde{\psi}_{\tilde{\mu}}:= \psi_{\mu(\tilde{\mu})}$ by
$\tilde{\mu}(\mu):= \int^{\mu} \md\nu/F(\nu)$. All derivative terms
can be shown to be corrections of higher order in $\hbar$, such that
one may use equidistant shift operators in $\tilde{\mu}$ up to quantum
corrections which one can absorb in the ordering choice
\cite{SchwarzN}. For specific refinement models of power-law form
$F(\mu)\propto \mu^x$, we have an equidistant equation in the variable
$\tilde{\mu}\propto\mu^{1-x}$, which would be proportional to volume
$\mu^{3/2}$ for $x=-1/2$.

In the region around $\mu\sim 0$, which is a strong quantum region,
the equidistant difference equation may differ significantly from the
original, non-equidistant one. Moreover, at those small volumes it is
difficult to find a specific form of refinement function $F(\mu)$
which could reliably capture refinements of small lattices. A power
law $f(p)\propto p^x$ with $x<0$ can clearly not be used there. It is
thus fortunate that the general difference equation (\ref{DiffEq}) can
be used to conclude general singularity avoidance: In a solution
scheme one can simply step over the values of the wave function at
$\mu=0$ where the classical singularity would be \cite{Sing,BSCG}. The
arguments given in \cite{Sing} for constant step-size are unchanged
for an equation of the form (\ref{DiffEq}) provided that $F(0)\not=0$
as it is required for a well-defined difference equation.  Thus,
quantum evolution of the wave function continues beyond the classical
singularity, which as this general statement has been shown also for
anisotropic \cite{HomCosmo,Spin} and spherically symmetric models
\cite{SphSymmSing}.

\subsection{Harmonic cosmology}

We are interested in quantum corrections as they arise when an initial
semiclassical state, which solves a difference equation such as
(\ref{DiffEq}) becomes more quantum. We can therefore make several
approximations by ignoring quantum effects which are not expected to
be strong or which would not change the qualitative effect of quantum
back-reaction. An example for the first effect is the inverse
$\langle\widehat{a^{-3}}\rangle$, which differs from $a^{-3}$ at small
scales. As such scales are not reached in what we consider
here,\footnote{Inverse scale factor corrections are not dominant in
isotropic models for large matter content (although they can be larger
than sometimes assumed; see the appendix of \cite{SchwarzN}). They do
become important in inhomogeneous situations where local energies
smaller than the total matter must be used.} we can assume
$\hat{a}^3\widehat{a^{-3}}\approx 1$. An example for the second effect
is factor ordering choices, which will change the precise values of
quantum correction terms but not their qualitative behavior. Thus, our
analysis is not sensitive to the fact that the ordering of the
Hamiltonian constraint in loop quantum cosmology is not fully fixed.
Taking a square root in (\ref{QuantFriedmann}), we thus obtain a
Schr\"odinger equation
\begin{equation} \label{Schroed}
 -\hat{p}_{\phi}\psi(p,\phi)=
i\hbar\frac{\partial}{\partial\phi}\psi(p,\phi)= \pm 
\sqrt{\frac{3}{4\pi G}}\gamma^{-1} 
\widehat{|f(p)^{-1}\sin(f(p)c)p|} \psi(p,\phi)
=: \pm\hat{H}\psi(p,\phi)
\end{equation}
whose main trace of the loop quantization is the occurrence of the
sine.\footnote{The harmless-looking absolute value on the right hand
side would be difficult to compute explicitly for operators, which
would make them non-local in specific $p$- or
$c$-representations. Here, however, this will not present a problem at
all: We simply need to require an initial state to have a
decomposition in $\hat{H}$-eigenstates where only states with positive
eigenvalues occur. Since $\hat{H}$ is preserved, this property will
continue to hold for evolved states. Imposing the absolute value is
thus not an issue of evolution, but of initial states which can be
much more easily dealt with (for $\Delta p_{\phi}\ll|p_{\phi}|$ which
we are going to use). One may also worry that superpositions are not
allowed of the right kind: positive- and negative frequency (i.e.\
$\hat{H}$-eigenstates with positive and negative eigenvalue,
respectively) instead of left- and right-moving. We do not allow
superpositions of the first kind as per the condition on initial
states. We do, however, allow superpositions of left- and right-moving
states (corresponding to the $\pm$-choice in (\ref{Schroed})). Note,
however, that we are going to compute primarily expectation values and
fluctuations, which are of interest only for individual wave packets
in a superposition, not for the total superposition.}

There is a specific factor ordering for which the system becomes
exactly solvable \cite{BouncePert}. Since there are no general
restrictions on the ordering, we choose this one for our
analysis. Again, we emphasize that conclusions drawn are reliable only
if they are insensitive to the factor ordering. For this, we use our
canonical variables $(f(p)c,V)$ and introduce $J:=V\exp(if(p)c)$ which
depends on $c$ only in a form allowed by the loop quantization. (Using
$V$ and $J$ as basic variables implies that the phase-space appears as
a cone since the mapping from $(p,c)$ to $(V,J)$ is not one-to-one at
$V=0$. But outside $V=0$, which is our main interest, we obtain a
correct description.) For operators, ordering $\hat{J}$ as indicated
in its definition, we have the commutation relations
\begin{equation}
  [\hat{V},\hat{J}]= \hbar\hat{J}\quad,\quad{}
  [\hat{V},\hat{J}^{\dagger}]=-\hbar\hat{J}^{\dagger} \quad,\quad{}
  [\hat{J},\hat{J}^{\dagger}]= -2\hbar\hat{V}-\hbar^2
\end{equation}
of an ${\rm sl}(2,{\mathbb R})$ algebra. Most importantly, the
Hamiltonian is a linear combination of these basic operators,
$\hat{H}=-\frac{1}{2}i(\hat{J}-\hat{J}^{\dagger})$, which guarantees
solvability, i.e.\ the decoupling of quantum variables from
expectation values.  (We are ignoring factors of $4\pi G/3$
and $1-x$ which would simply rescale $\phi$ in our solutions. They can
be reinstated simply by multiplying $H$ with $2\sqrt{4\pi G/3}(1-x)$,
which we will do in our effective Friedmann equation.)

We are still dealing with a difference equation of the type
(\ref{DiffEq}), but in a specific ordering which will allow us to get
access to interesting properties of its solutions much more easily via
expectation values and fluctuations. The specific difference equation
reads
\[
 -\hat{p}_{\phi}\psi = i\hbar\frac{\partial\psi}{\partial\phi}=
 \hat{H}\psi= \frac{1}{2i} (\tilde{\mu}\psi_{\tilde{\mu}-2}-
 (\tilde{\mu}+2)\psi_{\tilde{\mu}+2})
\]
which is a first-order version of
\[
 \hbar^2\frac{\partial^2\psi}{\partial\phi^2} = \frac{1}{4}
 (\tilde{\mu}(\tilde{\mu}-2) \psi_{\tilde{\mu}-4}-
 (\tilde{\mu}^2+(\tilde{\mu}+2)^2)\psi_{\tilde{\mu}}+
 (\tilde{\mu}+2)(\tilde{\mu}+4)\psi_{\tilde{\mu}+4})\,.
\]

Rather than solving for wave functions and then computing their
expectation values and fluctuations of $\hat{V}$ and $\hat{J}$, we
compute those quantities directly. Expectation values satisfy
equations of motion
\begin{equation} \label{expval}
 \frac{{\rm d}}{{\rm d}\phi}\langle\hat{V}\rangle =
 \frac{\langle[\hat{V},\hat{H}]\rangle}{i\hbar}= 
-\frac{1}{2}(\langle\hat{J}\rangle+\langle\hat{J}^{\dagger}\rangle)
\quad, \quad
 \frac{{\rm d}}{{\rm d}\phi}\langle\hat{J}\rangle =
 \frac{\langle[\hat{J},\hat{H}]\rangle}{i\hbar}=
-\langle\hat{V}\rangle-\frac{1}{2}\hbar=\frac{\md}{\md
 \phi}\langle\hat{J}^{\dagger}\rangle
\end{equation}
which are not coupled to fluctuations or higher moments and can be solved
easily. For arbitrary states, we have
\begin{eqnarray}
  \langle\hat{V}\rangle(\phi) &=& \frac{1}{2}(Ae^{-\phi}
  +Be^{\phi})-\frac{1}{2}\hbar \label{Vsol}\\
 \langle\hat{J}\rangle(\phi) &=& \frac{1}{2}(Ae^{-\phi}
 -Be^{\phi})+i H \label{Jsol}
\end{eqnarray}
with two constants of integration $A$ and $B$ and
$H=\langle\hat{H}\rangle$.

We have to impose reality conditions to ensure that the variable $c$
appearing in the complex $J$ is real. Classically, this is equivalent
to $J\bar{J}=V^2$, which also has to apply for operators:
$\hat{J}\hat{J}^{\dagger}= \hat{V}^2$. (But note
$\hat{J}^{\dagger}\hat{J}\not= \hat{V}^2$.) Upon taking an expectation
value, this condition relates expectation values and second order
quantum fluctuations:
\begin{equation} \label{reality}
 |\langle\hat{J}\rangle|^2-(\langle\hat{V}\rangle+
{\textstyle\frac{1}{2}}\hbar)^2=(\Delta V)^2-C_{J\bar{J}}+
 \frac{1}{4}\hbar^2
\end{equation}
where the $V$-fluctuation and $J$-$\bar{J}$-covariance
\begin{equation}
 (\Delta V)^2=\langle\hat{V}^2\rangle-\langle\hat{V}\rangle^2
 \quad\mbox{and}\quad C_{J\bar{J}}=
 \frac{1}{2}\langle\hat{J}\hat{J}^{\dagger}+
 \hat{J}^{\dagger}\hat{J}\rangle-
 \langle\hat{J}\rangle\langle\hat{J}^{\dagger}\rangle
\end{equation}
appear. If a state is semiclassical at least once, $(\Delta
V)^2-C_{J\bar{J}}$ is initially of the order $\hbar$ and remains so
because it is preserved in time \cite{BounceCohStates}. (See also the
explicit solutions (\ref{FlucSolI}) and (\ref{FlucSolVI}) below.)
Thus, inserting the solutions (\ref{Vsol}) and (\ref{Jsol}) yields
$AB=H^2+O(\hbar)$. This implies that $A$ and $B$ must have the same
sign, and we can define $A/B=:e^{2\delta}$ to write the solution as
$\langle\hat{V}\rangle(\phi)=H \cosh(\phi-\delta)$ (ignoring the small
contribution $-\frac{1}{2}\hbar$). Thus, there is no singularity of
vanishing volume but instead a bounce at $\phi=\delta$. Large $H$,
i.e.\ much matter in the universe, ensures that other quantum
corrections not included here do not destroy the bounce. For instance,
corrections in $\widehat{a^{-3}}$ or from the factor ordering are not
relevant in this case. For small $H$, however, there is no guarantee
for a semiclassical bounce as $(\Delta V)^2-C_{J\bar{J}}$ could
possibly compensate the positive $H^2$ and make $AB$ negative such
that $\langle\hat{V}\rangle(\phi)$ would be sinh-like. (There may be
quantizations in which all states bounce irrespective of whether they
are semiclassical even at only one time. But this would depend
sensitively on factor ordering choices, which we do not consider a
reliable property.)

\subsubsection{Free effective Friedmann equation}
\label{s:EffFriesFree}

As with any linear system, one can easily derive precise effective
equations for the harmonic model of cosmology. They have the advantage
of being more intuitive to interpret and thus give a more direct
handle on geometrical aspects which would be difficult to decipher
from a wave function. To derive effective equations
\cite{EffAc,Karpacz}, one first defines the quantum Hamiltonian as the
expectation value $H_Q=\langle\hat{H}\rangle$ in a general state,
parametrized by its expectation values and moments. The quantum
Hamiltonian is thus a function of the expectation values and all
infinitely many moments, determining their equations of motion via
Poisson brackets as they follow from quantum commutators. However, all
infinitely many equations of motion for the quantum variables are in
general coupled to each other and to the expectation values, which
usually makes these equations impossible to solve. One thus has to
look for suitable approximations based on a truncation of the coupled
system to a finite size of equations and variables. The truncated
equations are the effective equations, which describe properties of
the quantum system in regimes where the approximations used are valid.

In a linear system such as our solvable model for a quantum
space-time, no truncation is required as the variables automatically
decouple into sets of finitely many equations. Thus, the quantum
Hamiltonian provides precise effective equations which in our case
follow from $H_{\rm eff}= \langle\hat{H}\rangle= -\frac{1}{2}i
(J-\bar{J})$, only depending on expectation values. (In the context of
the effective Friedmann equation we drop brackets denoting expectation
values.) For expectation values, the equations (\ref{expval}) provided
before are thus precise effective equations. It is sometimes useful to
formulate them as evolution equations in proper time rather than
internal time $\phi$, as this can then readily be compared with the
classical Friedmann equation. To derive this, we use $H_{\rm
eff}=-i\sqrt{4\pi G/3}(1-x) (J-\bar{J})=p_{\phi}$, where we now have
reinstated the numerical factors, and write the equation of motion
\[
\frac{{\rm d}V}{{\rm d}\phi}= 
 -\sqrt{\frac{4\pi G}{3}}(1-x)(J+\bar{J})
=\mp 2\sqrt{\frac{4\pi G}{3}}(1-x)V\sqrt{1-\frac{3p_{\phi}^2}{16\pi G(1-x)^2V^2}+
\sigma}\,.
\]
Here, we have eliminated $J+\bar{J}$ using the reality condition
(\ref{reality}) which implies
\[
 \frac{1}{4}\left((J+\bar{J})^2+(i(J-\bar{J}))^2\right)= J\bar{J}
 = (V+\hbar/2)^2+(\Delta
 V)^2-G^{J\bar{J}}+\frac{\hbar^2}{4}
\]
and thus
\begin{equation} \label{trig}
 \frac{J+\bar{J}}{2V+\hbar}= \pm\sqrt{1-\left(\frac{J-\bar{J}}{i(2V+\hbar)}
\right)^2+   \sigma}
\end{equation}
with a small $\sigma:=((\Delta
V)^2-G^{J\bar{J}}+\hbar^2/4)/(V+\hbar/2)^2$ as a relative quantum
variables. (This relation reflects the identity
$\cos(f(p)c)^2+\sin(f(p)c)^2=1$, corrected by factor ordering terms
which arise from the quantized $J$. The contribution
$\frac{1}{2}\hbar$ to $V$ can safely be ignored for our purposes.
Notice that the derivation of $\sigma$ shows that its form depends on
factor ordering choices in the Hamiltonian and basic variables.)

To reformulate this as a Friedmann equation, we use the expression for
$V$ in terms of $a$ of (\ref{V}) for a refinement function
$f(p)=f_0p^x$.  Proper time then enters through\footnote{As we are not
including quantum corrections in the matter Hamiltonian, this relation
is identical to the classical one.}  $\dot{\phi}=\{\phi,H_{\rm
matter}\}=a^{-3}p_{\phi}$, which implies the corrected Friedmann
equation
\begin{equation} \label{rhosquared} 
\left(\frac{\dot{a}}{a}\right)^2=
  \left(\frac{1}{2(1-x)}\right)^2 \left(\frac{\dot{V}}{V}\right)^2
  = \frac{4\pi G}{3} \frac{p_{\phi}^2}{a^6} \left(1-\frac{4\pi G}{3}
  \gamma^2 f_0^2 a^{2+4x} \frac{p_{\phi}^2}{a^6}\right)
\end{equation}
for a free, massless scalar.  As one can easily see, the quantum
correction can simply be formulated as a term quadratic in the energy
density of the free scalar \cite{RSLoopDual} (see also
\cite{EffHam,GenericBounce,AmbigConstr}).  The bounce manifests itself
since $\dot{a}=0$ is possible if $\rho_{\rm
free}=\frac{1}{2}p_{\phi}^2/a^6= \rho_{\rm crit}=3 a^{-2-4x}/8\pi G
f_0^2\gamma^2$. For $x=-1/2$ and $f_0\sim\lP$, this is a Planckian
energy density, but can be smaller\footnote{Moreover, the critical
density would be $a$-dependent for $x\not=-1/2$. In this case, it is
not just the matter density which plays a role for the bounce but also
the underlying quantum gravitational state and its refinement. There
is no a priori reason why only the matter density should play a role
for the bounce. In fact, it is the spatial discreteness of the quantum
representation which implies a repulsive force, and thus the quantum
gravity state is the primary reason for the bounce. In the language of
refinement models, including the coordinate volume $V_0$, we have
$f_0^2a^{2+4x}=f(p)^2p\sim {\cal N}^{-2/3}p=(a^3/{\cal N})^{2/3}$ which
is the value of an elementary flux, or roughly the area $L^2$ of an
elementary lattice plaquette. In terms of $L^2$, we have the critical
density $\rho_{\rm crit}= 3/8\pi G\gamma^2L^2$ for any $x$, which
clearly shows that the critical density is determined by the
underlying lattice. For $x=-1/2$, $L$ happens to be independent of
$a$, and thus the critical density is a constant. For $x\not=-1/2$, on
the other hand, the elementary flux depends on the total spatial size
in a way determined by the lattice refinement, thus also making the
critical density $a$-dependent. Requiring $L$ to be constant and of
the order $\lP$ implies that $x=-1/2$ and makes the critical density
Planckian, which produces the parameter choices of
\cite{APS}. However, this requirement is not necessary.} for larger
$f_0$ or $x>-1/2$.

For a general interpretation, one should note, however, that higher
powers of $\rho$ are not the primary correction in loop quantum
cosmology. The reason for the new term is the higher curvature
corrections from the presence of the sine in the effective
Hamiltonian, which in this model can be reformulated as a simple
correction of the energy dependence. This proves the correctness of
the effective Friedmann equation, as realized in \cite{BouncePert},
but only for this specific matter content.  While it is tempting to
use (\ref{rhosquared}) for all matter contents, just replacing
$p_{\phi}^2/2a^6$ with $\rho$, this would overlook corrections arising
from quantum back-reaction which must be present in general. We will
come back to these corrections in the next section, where we derive
correct effective equations in the presence of self-interactions.

\subsubsection{Quantum variables}

The solution procedure can be repeated for higher moments of a state,
for which the same effective Hamiltonian $H_{\rm eff}=
-\frac{1}{2}i(\langle\hat{J}\rangle- \langle\hat{J}^{\dagger}\rangle)$
provides equations of motion.  Hamiltonian equations of motion, for
instance for volume fluctuations $(\Delta
V)^2=\langle\hat{V}^2\rangle-\langle\hat{V}\rangle^2$, are then
derived by means of Poisson brackets which follow from expectation
values of commutators: using
$\{\langle\hat{V}^2\rangle,\langle\hat{J}\rangle\}=
\langle[\hat{V}^2,\hat{J}]\rangle/i\hbar=
-i\langle\hat{V}\hat{J}+\hat{J}\hat{V}\rangle$ and
$\{\langle\hat{V}\rangle^2,\langle\hat{J}\rangle\}=
2\langle\hat{V}\rangle \langle[\hat{V},\hat{J}]\rangle/i\hbar= -2i
\langle\hat{V}\rangle \langle\hat{J}\rangle$ we have $\{(\Delta
V)^2,\langle\hat{J}\rangle\}= -2i C_{VJ}$. Other Poisson brackets can
be derived analogously (see the App.~\ref{a:Poisson} and
\cite{BouncePot} for explicit expressions), which results in equations
of motion
\begin{eqnarray}
 \frac{\md}{\md\phi}(\Delta V)^2 &=& -C_{VJ}-C_{V\bar{J}}\quad,\quad
 \frac{\md}{\md\phi}(\Delta J)^2 = -2C_{VJ}\quad,\quad
 \frac{\md}{\md\phi}(\Delta \bar{J})^2 = -2C_{V\bar{J}}\nonumber\\ 
 \frac{\md}{\md\phi}C_{VJ} &=&
 -\frac{1}{2} (\Delta J)^2-\frac{1}{2}C_{J\bar{J}}-(\Delta V)^2\quad,\quad
 \frac{\md}{\md\phi}C_{V\bar{J}} = -\frac{1}{2} (\Delta \bar{J})^2-
 \frac{1}{2}C_{J\bar{J}}-(\Delta V)^2\nonumber\\ 
 \frac{\md}{\md\phi}C_{J\bar{J}} &=&
 -C_{VJ}-C_{V\bar{J}}\,.
\end{eqnarray}

They are solved by \cite{BounceCohStates}
\begin{eqnarray}
 (\Delta V)^2(\phi) &=& \frac{1}{2}(c_3e^{-2\phi}+c_4e^{2\phi})- 
\frac{1}{4}(c_1+c_2)\label{FlucSolI}\\
 (\Delta J)^2(\phi) &=& \frac{1}{2}(c_3e^{-2\phi}+c_4e^{2\phi})+
\frac{1}{4}(3c_2-c_1)-
 i(c_5e^{\phi}-c_6e^{-\phi})\\
 (\Delta \bar{J})^2(\phi) &=& \frac{1}{2}(c_3e^{-2\phi}+c_4e^{2\phi})+
\frac{1}{4}(3c_2-c_1)+
 i(c_5e^{\phi}-c_6e^{-\phi})\\
 C_{VJ}(\phi) &=& \frac{1}{2}(c_3e^{-2\phi}-c_4e^{2\phi})+ 
\frac{i}{2}(c_5e^{\phi}+c_6e^{-\phi})\\
 C_{V\bar{J}}(\phi) &=& \frac{1}{2}(c_3e^{-2\phi}-c_4e^{2\phi})- 
\frac{i}{2}(c_5e^{\phi}+c_6e^{-\phi})\\
 C_{J\bar{J}}(\phi) &=& \frac{1}{2}(c_3e^{-2\phi}+c_4e^{2\phi})+
 \frac{1}{4}(3c_1-c_2)\label{FlucSolVI}
\end{eqnarray}
with independent integration constants $c_1,\ldots,c_6$. Reality
conditions are already implemented for real $c_I$ since this implies
$C_{V\bar{J}}= \overline{C_{VJ}}$ and\footnote{Note that it is not the
square root $\Delta\bar{J}$ for which reality conditions are imposed
because they are based on $\langle(\hat{J}^{\dagger})^2\rangle-
\langle\hat{J}^{\dagger}\rangle^2=
\overline{\langle(\hat{J})^2\rangle- \langle\hat{J}\rangle^2}$.}
$(\Delta \bar{J})^2= \overline{(\Delta J)^2}$. Solutions can be chosen
to saturate uncertainty relations in order to determine properties of
dynamical coherent states. This analysis has been performed in
\cite{BounceCohStates,Harmonic} in order to shed light on the possible
asymmetries of volume fluctuations before and after the bounce.  For
large $H$, which is required for a massive universe, state properties
such as $c_3/c_4$ and $c_5/c_6$ are extremely sensitive to initial
values.  Thus, from current knowledge of the classicality of the
universe one can practically derive nothing about the precise
fluctuations of the state before the big bang. As we will see below,
these same parameters are also crucial for the behavior near the
would-be bounce at $\phi=0$ of an interacting system. We emphasize the
role of $H$: a large $H$ keeps the bounce away from small volume and
its strong quantum effects, but heightens the sensitivity to small
changes in an initial quantum state. Small $H$ would reduce the
sensitivity, but also the bounce volume and thus introduce more
quantum effects in the evolution.  When discussing interactions in
what follows we will see further implications of a large value of $H$.

\subsection{Interactions}
\label{s:Inter}

We are now ready to introduce an interaction term via a potential
$W(\phi)$ of the scalar, which destroys the exact
solvability and implies quantum back-reaction. There are several
difficulties, which complicate a complete analysis. All this has been
discussed in \cite{BouncePot}, and we present a brief review here:
First, $\phi$ in general will no longer serve as a global internal
time, such that one can only study patches of solutions in which
$\phi$ would be a monotonic function of a coordinate time. Especially
since we have to address a question about long evolution times,
several patches are in general required. Secondly, the
$\phi$-Hamiltonian $p_{\phi}$ is now time dependent, as it is a
function of $\phi$ through the potential. While
\begin{equation} \label{pphipot}
-p_{\phi}=\pm H= \pm \frac{3}{4\pi G}|p| \sqrt{\frac{c^2}{\gamma^2}-
\frac{8\pi G}{3}|p| W(\phi)}
\end{equation}
can still be used as a classical Hamiltonian, upon quantization
solutions to the equation
$-\hat{p}_{\phi}\psi=i\hbar\partial\psi/\partial\phi=\pm\hat{H}\psi$
will no longer provide exact solutions of the original constraint
equation (\ref{QuantFriedmann}) which is a second order differential
equation in $\phi$: for solutions $\psi$ of the linear equation, we
have $\hat{p}_{\phi}^2\psi=\pm\hat{p}_{\phi}\hat{H}\psi=
\hat{H}^2\psi\pm[\hat{p}_{\phi},\hat{H}]\psi
\not=\hat{H}^2\psi$. Finally, there are now couplings between all the
quantum variables, which requires a detailed analysis of effective
equations.

For a sufficiently small and flat potential, all these difficulties
can be evaded: (i) While there is no global time, a small flat
potential allows long monotonic changes of $\phi$ which one can
analyze in one patch. Different patches can be combined, but this
requires the consideration of general states as initial semiclassical
states in the first patch will change in each patch and possibly in
the patching process. (ii) The commutator $[\hat{p}_{\phi},\hat{H}]$
which determines the error one makes by formulating the initial second
order equation (\ref{QuantFriedmann}) as a first order Schr\"odinger
equation (\ref{Schroed}) is proportional to $\hbar W'$ and thus small
for a flat potential. Moreover, since it arises from a commutator, one
may view this extra term as simply a quantization ambiguity, or
correct it by adding an extra contribution to the Hamiltonian
$\hat{H}$ used for the linear equation. (iii) Coupling terms can be
handled by suitable approximations in effective equations. This will
be the main focus here.

\subsubsection{Effective equations}

To analyze implications for the bounce, we have to change the
classical Hamiltonian (\ref{pphipot}) in two ways: we use canonical
variables for general $x$, taking into account different refinement
schemes of loop quantum cosmology, and then introduce the holonomy
variable $J$ and its complex conjugate. This gives (again dropping,
for now, numerical factors)
\begin{equation}
 H=\sqrt{-\frac{1}{4}(J-\bar{J})^2- V^{3/(1-x)} W(\phi)}=
 \frac{J-\bar{J}}{2i}- i\frac{V^{3/(1-x)}}{J-\bar{J}} W(\phi)+\cdots
\end{equation}
in an expansion by the potential. For this non-linear classical
Hamiltonian, the quantum Hamiltonian depends on all quantum variables.
It can be derived from a formal Taylor expansion of the expression
$H(\langle\hat{V}\rangle+(\hat{V}-\langle\hat{V}\rangle),
\langle\hat{J}\rangle+(\hat{J}-\langle\hat{J}\rangle),
\langle\hat{J}^{\dagger}\rangle+ (\hat{J}^{\dagger}-
\langle\hat{J}^{\dagger}\rangle))$ in $\hat{V}-\langle\hat{V}\rangle$,
$\hat{J}-\langle\hat{J}\rangle$ and $\hat{J}^{\dagger}-
\langle\hat{J}^{\dagger}\rangle$. From the result one can derive
Hamiltonian equations of motion via Poisson brackets between
expectation values and quantum variables, a formal procedure which can
be seen as a shortcut to computing expectation values of commutators
$\langle[\cdot,\hat{H}]\rangle$.

In our case, we obtain the quantum Hamiltonian
\begin{eqnarray} \label{HQ}
 H_Q &=& \frac{J-\bar{J}}{2i}- i\frac{V^{3/(1-x)}}{J-\bar{J}}
 W(\phi)\\
 && - \frac{3}{2}i\frac{2+x}{(1-x)^2} \frac{V^{(1+2x)(1-x)}}{J-\bar{J}}
 (\Delta V)^2W(\phi)+ \frac{3i}{1-x} \frac{V^{(2+x)/(1-x)}}{(J-\bar{J})^2}
 (C_{VJ}-C_{V\bar{J}}) W(\phi)\nonumber\\
 &&- i\frac{V^{3/(1-x)}}{(J-\bar{J})^3} ((\Delta J)^2
 -2C_{J\bar{J}}+(\Delta \bar{J})^2)W(\phi)+\cdots \nonumber
\end{eqnarray}
where the omitted terms now also include higher moments. We will
regard this truncation as our perturbative effective Hamiltonian,
which is valid as long as higher moment terms are subdominant.

Due to the new coupling terms, equations of motion become lengthy and
we do not present all of them here. Complete equations to the order
used here and details of their derivation can be found in
\cite{BouncePot} for $x=0$.  The main equation of interest here is
that for $\langle\hat{V}\rangle$ because it is the one giving us
information about the bounce.  From the effective Hamiltonian, we find
its equation of motion
\begin{eqnarray} \label{Vdotpert}
\frac{{\rm d}\langle\hat{V}\rangle}{{\rm d}\phi} &=& 
-\frac{\langle\hat{J}\rangle+\langle\hat{J}^{\dagger}\rangle}{2}+
\frac{\langle\hat{J}\rangle+\langle\hat{J}^{\dagger}\rangle}%
{(\langle\hat{J}\rangle-\langle\hat{J}^{\dagger}\rangle)^2} 
\langle\hat{V}\rangle^{3/(1-x)} W(\phi)\\
&&+3  \frac{2+x}{(1-x)^2}\frac{\langle\hat{J}\rangle+
\langle\hat{J}^{\dagger}\rangle}%
{(\langle\hat{J}\rangle-\langle\hat{J}^{\dagger}\rangle)^2}
\langle\hat{V}\rangle^{(1+2x)/(1-x)} (\Delta V)^2\,W(\phi)
\nonumber\\
&&+ 3
\frac{\langle\hat{J}\rangle+\langle\hat{J}^{\dagger}\rangle}%
{(\langle\hat{J}\rangle-\langle\hat{J}^{\dagger}\rangle)^4}
\langle\hat{V}\rangle^{3/(1-x)}
((\Delta J)^2+(\Delta \bar{J})^2-2C_{J\bar{J}})W(\phi)
 \nonumber\\
&& -\frac{6}{1-x} \frac{\langle\hat{J}\rangle+\langle\hat{J}^{\dagger}\rangle}%
{(\langle\hat{J}\rangle-\langle\hat{J}^{\dagger}\rangle)^3}
\langle\hat{V}\rangle^{(2+x)/(1-x)} 
(C_{VJ}- C_{V\bar{J}})W(\phi)\nonumber\\
&&- \frac{2\langle\hat{V}\rangle^{3/(1-x)}}{(\langle\hat{J}\rangle-
\langle\hat{J}^{\dagger}\rangle)^3} 
((\Delta J)^2-(\Delta \bar{J})^2)W(\phi)\nonumber\\
&&+\frac{3}{1-x}\frac{\langle\hat{V}\rangle^{(2+x)/(1-x)}}{(\langle
\hat{J}\rangle-
\langle\hat{J}^{\dagger}\rangle)^2} 
(C_{VJ}+C_{V\bar{J}})W(\phi)\,.\nonumber
\end{eqnarray}
The evolution of the volume expectation value now depends on the
behavior of the state via its moments. 

For semiclassical states, the correction terms are certainly small,
but we recall that the singularity problem is a question about
long-term evolution. Then, the state may change considerably and our
equation, corresponding to the last phase before the big bang is
reached, has to be applied to a state whose moments can be large.
Additional correction terms will arise from higher moments neglected
so far, but we can use the equation to test whether the fluctuations
have an effect, and how this can occur precisely. Moreover, our
conclusions can be extended to all orders in the potential and in
quantum variables, which allows for arbitrary states
\cite{BounceSqueezed}. We are here not interested in a precise
analysis, which would seem premature, but in understanding the role of
quantum corrections. Thus, we assume them to be small and check the
self-consistency of this assumption.

\subsubsection{Correlations}

Let us thus assume that our state has for some time evolved by the
exact solutions of the solvable model, corrected only slightly by the
classical potential term and quantum back-reaction. We may thus use
the solutions (\ref{Vsol}) and (\ref{Jsol}) for expectation values and
(\ref{FlucSolI})--(\ref{FlucSolVI}) for fluctuations and covariances
as zeroth order solutions to estimate the magnitude of quantum
back-reaction. In particular, we look at where the free solutions
would have their bounce, thus implying
$\langle\hat{J}\rangle+\langle\hat{J}^{\dagger}\rangle\approx 0$. This
already eliminates several of the terms in (\ref{Vdotpert}), in
particular the classical interaction. However, the last two quantum
correction terms remain, which further using
$\langle\hat{V}\rangle\approx H\approx
-\frac{1}{2}i(\langle\hat{J}\rangle-\langle\hat{J}^{\dagger}\rangle)$
gives
\begin{equation} \label{VphiSq}
 \frac{{\rm d}\langle\hat{V}\rangle}{{\rm d}\phi} \approx
 -\frac{1}{H^{-3x/(1-x)}}\left(\frac{1}{2}{\rm Im}(\Delta J(\phi))^2+
 \frac{3}{2(1-x)}{\rm Re} C_{VJ}(\phi)\right)W(\phi)\,.
\end{equation}
This vanishes around $\phi\approx0$ for background solutions
satisfying $c_3\sim c_4$, $c_5\sim c_6$, which corresponds to
unsqueezed states. For squeezed states, however, the time derivative
of $\langle\hat{V}\rangle$ does not vanish for the interacting system
at the bounce of the free solutions. The interacting system cannot
bounce where the free system would bounce. {\em Thus, quantum
back-reaction does affect the bounce for squeezed states.} Whether or
not and where the bounce may happen depends on quantum properties of
the state, not just on expectation values as in the free case. Even
for free solutions, there is not much control over the key parameters
$c_3/c_4$ and $c_5/c_6$ due to cosmic forgetfulness: it is precisely
these parameters which determine the asymmetry of fluctuations
discussed in \cite{BeforeBB,Harmonic}. It will thus be even more
complicated to constrain these parameters in an interacting system.
To estimate the precise effect we have to know how squeezing develops
for the interacting solutions; otherwise no reliable statement about
the persistence of the bounce in the perturbed system can be made.

In addition to cosmic forgetfulness, we have to consider the dynamical
form of quantum variables and their coupling to expectation values in
interacting states. Notice that the quantum corrections in
(\ref{Vdotpert}) split into two classes: the first three lines which
include $\langle\hat{J}\rangle+ \langle\hat{J}^{\dagger}\rangle$ as a
factor and which can be subsumed as changing the classical potential
to $W(\phi)(1+\epsilon_1)$ with relative fluctuations
\begin{equation} \label{eps}
 \epsilon_1= 3\frac{2+x}{(1-x)^2}\frac{(\Delta V)^2}{\langle\hat{V}\rangle^2}+
3\frac{(\Delta J)^2-2C_{J\bar{J}}+
\Delta\bar{J})^2}{(\langle\hat{J}\rangle-
(\langle\hat{J}^{\dagger}\rangle)^2}- \frac{6}{1-x}
\frac{C_{VJ}-C_{V\bar{J}}}{\langle\hat{V}\rangle
  (\langle\hat{J}\rangle- \langle\hat{J}^{\dagger}\rangle)}\,,
\end{equation}
and the last two lines which are present even when
$\langle\hat{J}\rangle+ \langle\hat{J}^{\dagger}\rangle$
vanishes. 

These last terms are of special interest because they remain in the
analysis of the bounce which led to (\ref{VphiSq}). To see whether
they are generically zero or, if not, how large they can grow, we need
the equations of motion for $C_{VJ}$ and $(\Delta J)^2$. This requires
more Poisson brackets than presented so far, which are listed in
App.~\ref{a:Poisson} as recalled from \cite{BouncePot}.  In these
relations, moments of third order appear (with triple superscript
indices) which we ignore in our approximation together with the terms
of order $\hbar^2$ or higher.

With these relations and the quantum Hamiltonian (\ref{HQ}), we obtain
\begin{eqnarray}
 \frac{\md}{\md\phi}({\rm Re}C_{VJ}) &=& -\frac{1}{2} ({\rm Re}(\Delta
 J)^2+ C_{J\bar{J}}+ 2(\Delta V)^2) \\
&& +2\frac{\langle\hat{V}\rangle^{3/(1-x)}}{(\langle\hat{J}\rangle-
   \langle\hat{J}^{\dagger}\rangle)^2}
 \left(C_{J\bar{J}}+\frac{10-x}{1-x} (\Delta V)^2\right)
 W(\phi)\nonumber\\
&&+ \frac{3}{2}\frac{2+x}{(1-x)^2}
 \langle\hat{V}\rangle^{(1+2x)/(1-x)} (\Delta V)^2 W(\phi)\nonumber\\
 && + \left(\frac{6i}{1-x}
   \frac{\langle\hat{V}\rangle^{(2+x)/(1-x)}}{\langle\hat{J}\rangle-
     \langle\hat{J}^{\dagger}\rangle}- 12i
   \frac{\langle\hat{V}\rangle^{(4-x)/(1-x)}}{(\langle\hat{J}\rangle-
     \langle\hat{J}^{\dagger}\rangle)^3}\right) {\rm Im}C_{VJ}
 W(\phi)\nonumber\\
&&- \frac{3}{1-x} \frac{\langle\hat{V}\rangle^{(2+x)/(1-x)}
   (\langle\hat{J}\rangle+
   \langle\hat{J}^{\dagger}\rangle)}{(\langle\hat{J}\rangle-
   \langle\hat{J}^{\dagger}\rangle)^2} {\rm Re}C_{VJ} W(\phi)\nonumber\\
 && -i\frac{\langle\hat{V}\rangle^{3/(1-x)} (\langle\hat{J}\rangle+
   \langle\hat{J}^{\dagger}\rangle)}{(\langle\hat{J}\rangle-
   \langle\hat{J}^{\dagger}\rangle)^3} {\rm Im}(\Delta J)^2 W(\phi)\nonumber
\end{eqnarray}
and
\begin{eqnarray}
 \frac{\md}{\md\phi} ({\rm Im} (\Delta J)^2) &=& -2{\rm Im}C_{VJ}
 +4\left(\frac{5+x}{1-x}
   \frac{\langle\hat{V}\rangle^{3/(1-x)}}{(\langle\hat{J}\rangle-
     \langle\hat{J}^{\dagger}\rangle)^2}- \frac{3(2+x)}{2(1-x)^2}
   \langle\hat{V}\rangle^{(1+2x)/(1-x)}\right) {\rm Im}C_{VJ}
 W(\phi)\nonumber\\
&& -2i\left(3\frac{2+x}{(1-x)^2}-4
  \frac{\langle\hat{V}\rangle^2}{(\langle\hat{J}\rangle-
    \langle\hat{J}^{\dagger}\rangle)^2} \right)
\frac{\langle\hat{V}\rangle^{(1+2x)/(1-x)} (\langle\hat{J}\rangle+
  \langle\hat{J}^{\dagger}\rangle)}{\langle\hat{J}\rangle-
  \langle\hat{J}^{\dagger}\rangle} {\rm Re}C_{VJ} W(\phi)\nonumber\\
&&-8i\frac{\langle\hat{V}\rangle^{(4-x)/(1-x)}}{(\langle\hat{J}\rangle-
    \langle\hat{J}^{\dagger}\rangle)^3} (2(\Delta V)^2+C_{J\bar{J}})
  W(\phi)\nonumber\\
&& -2i\left(\frac{3}{1-x}-
  4\frac{\langle\hat{V}\rangle^2}{(\langle\hat{J}\rangle-
    \langle\hat{J}^{\dagger}\rangle)^2} \right)
\frac{\langle\hat{V}\rangle^{(2+x)/(1-x)}}{\langle\hat{J}\rangle-
  \langle\hat{J}^{\dagger}\rangle} {\rm Re}(\Delta J)^2 W(\phi)\nonumber\\
&&-\frac{6}{1-x}
\frac{\langle\hat{V}\rangle^{(2+x)/(1-x)}(\langle\hat{J}\rangle+
  \langle\hat{J}^{\dagger}\rangle)}{(\langle\hat{J}\rangle-
  \langle\hat{J}^{\dagger}\rangle)^2} {\rm Im}(\Delta J)^2 W(\phi)\,.
\end{eqnarray}

Derivatives of the correlations by $\phi$ cannot vanish because they
are determined by some of the fluctuations which, thanks to
uncertainty relations, are non-zero. For instance, $(\Delta V)^2$
appears in both equations and drives the change of squeezing. It is
multiplied by the small potential, but also by factors of the volume.
If we consider relative fluctuations $(\Delta V)^2/V^2$ of volume,
which are nearly constant in each expanding or contracting branch of
the free solutions, the dominating pre-factors are
$\langle\hat{V}\rangle^{(5-2x)/(1-x)}/(\langle\hat{J}\rangle-
\langle\hat{J}^{\dagger}\rangle)^2$ for ${\rm Re}C_{VJ}$ and
$\langle\hat{V}\rangle^{3(2-x)/(1-x)}/(\langle\hat{J}\rangle-
\langle\hat{J}^{\dagger}\rangle)^3$ for ${\rm Im}(\Delta J)^2$. Thus,
especially at large volume, but also at the bounce itself, these terms
can contribute to a significant change in time of squeezing. For
$x=-1/2$, for instance, they behave as $V^4/H^2$ and $V^5/H^3$,
respectively. For the expanded Hamiltonian, on the other hand, only
$V^2/H^2W\ll 1$ is required, which allows the pre-factors to be
large. Moreover, since the arguments can be extended to all orders in
powers of the potential \cite{BounceSqueezed}, the condition on $W$ to
be small can be relaxed. When small volume is reached, squeezing
parameters will be important in Eq.~(\ref{VphiSq}).

Eq.~(\ref{VphiSq}) then shows that a large $H$ has a protective effect
for the bounce for $x\not=0$, which is strongest for the limiting case
$x=-1/2$. As before, we thus notice that a large $H$, which implies a
bounce at large volume, preserves classicality better than small $H$,
which is more pronounced the larger $-x$ is. However, the influence of
squeezing is always present and is required for a proper understanding
of the bounce. Also this is weaker for $x=-1/2$ than for $x=0$, but
remains potentially large. We will now analyze the same effect from
the point of view of an effective Friedmann equation which, unlike
(\ref{rhosquared}), correctly captures state properties in the
presence of interactions.

\subsubsection{Effective Friedmann equation}

Using the effective Hamiltonian, which is the truncation of the
quantum Hamiltonian explicitly written out in (\ref{HQ}), and the
$\phi$-evolution equation (\ref{Vdotpert}) of $\langle\hat{V}\rangle$,
we can derive an effective Friedmann equation as before in
Sec.~\ref{s:EffFriesFree}. There are now several additional terms
especially due to quantum variables, whose explicit dynamics we have
not determined. We will therefore keep them in a general form as
relative fluctuations such as (\ref{eps}) and see how they affect the
Friedmann equation.

We first sketch the procedure and non-trivial changes due to the
potential for the expanded Hamiltonian linear in $W$, but then use
also the quadratic terms shown in App.~\ref{a:second} in the final
effective Friedmann equation because they are expected if
$\rho^2$-corrections are present.  In the effective Hamiltonian, it
turns out that all relative quantum variables can be summarized by a
single multiplicative correction to the classical potential:
\[
 H_{\rm eff}= \frac{J-\bar{J}}{2i}
-i\frac{V^{3/(1-x)}}{J-\bar{J}} W(\phi)(1+\delta_1)
\]
with
\begin{equation} \label{delta1}
  \delta_1= \frac{3}{2}\frac{2+x}{(1-x)^2}
\frac{(\Delta V)^2}{\langle\hat{V}\rangle^2}+
\frac{(\Delta J)^2-2C_{J\bar{J}}+
(\Delta\bar{J})^2}{(\langle\hat{J}\rangle-
(\langle\hat{J}^{\dagger}\rangle)^2)}- \frac{3}{1-x}
\frac{C_{VJ}-C_{V\bar{J}}}{\langle\hat{V}\rangle
  (\langle\hat{J}\rangle- \langle\hat{J}^{\dagger}\rangle)}
\end{equation}
which differs from $\epsilon_1$ in (\ref{eps}) by some
coefficients. 

To express $\md V/\md\phi$ in a similar form, we have to distinguish
between two different contributions from quantum variables. One of
them is $\epsilon_1$, which, like $\delta_1$, just multiplies the
potential. But there are additional contributions in the last two
lines of (\ref{Vdotpert}) containing squeezing parameters, which are
not of this form and which we write as\footnote{There is some
arbitrariness in what one considers as relative fluctuations. Had we
used $\langle\hat{J}\rangle+ \langle\hat{J}^{\dagger}\rangle$ instead
of what appears in $\eta_1$, then there would be a total factor of
$\langle\hat{J}\rangle+\langle\hat{J}^{\dagger}\rangle$ and $\eta_1$
could be subsumed in $\epsilon_1$. It might, in fact, appear more
natural to use, e.g.,
$(C_{VJ}+C_{V\bar{J}})/\langle\hat{V}\rangle(\langle\hat{J}\rangle+
\langle\hat{J}^{\dagger}\rangle)$ instead of
$(C_{VJ}+C_{V\bar{J}})/\langle\hat{V}\rangle^2$. However, it is
dangerous to do that because one would divide by a quantity,
$\langle\hat{J}\rangle+ \langle\hat{J}^{\dagger}\rangle$, which then
vanishes at the bounce. The quantity $\eta_1$, on the other hand,
simply absorbs factors which are already present in the equation or
which do not vanish, and does so in a way which makes the final
quantity dimensionless. Moreover, in each expanding or contracting
branch $\langle\hat{V}\rangle$ behaves approximately as
$\langle\hat{J}\rangle+\langle\hat{J}^{\dagger}\rangle$ but does not
become zero. All relative quantum variables defined here are nearly
constant for the free solutions in each branch, although they can
change during the free bounce.}
\begin{equation} \label{eta1}
 \eta_1 = -2 \frac{(\Delta J)^2-
   (\Delta\bar{J})^2}{\langle\hat{V}\rangle(\langle\hat{J}\rangle-
\langle\hat{J}^{\dagger}\rangle)}+ \frac{3}{1-x}
\frac{C_{VJ}+C_{V\bar{J}}}{\langle\hat{V}\rangle^2}\,.
\end{equation}
We then have the evolution equation
\begin{equation} \label{dVdphiint}
 \frac{\md V}{\md\phi} = -\frac{J+\bar{J}}{2}+
 \frac{J+\bar{J}}{(J-\bar{J})^2} V^{3/(1-x)} W(\phi)(1+\epsilon_1) +
 \frac{V^{(4-x)/(1-x)}}{(J-\bar{J})^2} \eta_1 W(\phi)\,.
\end{equation}

With all numerical factors reinstated, we have
\[
 H_{\rm
eff}=2\sqrt{\frac{4\pi G}{3}} (1-x) \left(\frac{J-\bar{J}}{2i}- i
\frac{8\pi G}{3} \left(\frac{8\pi\gamma Gf_0(1-x)V}{3}\right)^{3/(1-x)}
 \frac{W(\phi)}{J-\bar{J}}
 (1+\delta_1)\right)=p_{\phi}
\]
which provides a quadratic equation for $J-\bar{J}$ in terms of
$p_{\phi}$. Its perturbative solution in $W$ and fluctuations is
\[
 2i(J-\bar{J}) = \sqrt{\frac{3}{4\pi G}} \frac{p_{\phi}}{2(1-x)} 
\left(1+a^6\frac{W(\phi)}{p_{\phi}^2} (1+\delta_1)\right)
\]
and, via (\ref{trig}), provides
\[
 \frac{J+\bar{J}}{2V+\hbar} = \pm\sqrt{1-\frac{8\pi\gamma^2Gf_0^2}{3}
 a^{2+4x}\rho+\sigma+O\left((a^6Wp_{\phi}^{-2})^2\right)} 
\]
with the energy density $\rho=\frac{1}{2}a^{-6}p_{\phi}^2+W(\phi)$ of
the interacting scalar.

The evolution equation (\ref{dVdphiint}) now is to be squared for the
effective Friedmann equation after replacing $J+\bar{J}$. Here, terms
quadratic in the potential become important, and so we should include
them in the quantum Hamiltonian expanded to second order in the
potential. The complete expressions are provided in
App.~\ref{a:second}, which, with all numerical factors reinstated,
provides the effective Friedmann equation
\begin{eqnarray} \label{EffFriedQBR}
 \left(\frac{\dot{a}}{a}\right)^2 &=& \frac{8\pi G}{3}
 \left(\rho+\epsilon_1 W(\phi)- 2\xi_2 a^6
 \frac{W^2}{p_{\phi}^2}\right)\nonumber\\
 &&\qquad\times \left(1-\frac{8\pi G}{3}\gamma^2 f_0^2 a^{2+4x}
 \left(\rho+W(\phi)\delta_1+ 
\frac{\xi_1}{2} a^6
 \frac{W^2}{p_{\phi}^2}\right)+\sigma \right)
\nonumber\\
 && \pm\frac{4\pi G}{3} W(\phi) 
\sqrt{1-\frac{8\pi G}{3}\gamma^2 f_0^2 a^{2+4x}
 \left(\rho+W(\phi)\delta_1+ 
\frac{\xi_1}{2} a^6
 \frac{W^2}{p_{\phi}^2}\right)+\sigma}\nonumber\\
&&\qquad\times\left(\eta_1+\xi_3\frac{a^6W(\phi)}{p_{\phi}^2}\right)\nonumber\\
 && + \frac{4\pi G}{3}\frac{a^{6} W(\phi)^2}{p_{\phi}^2} \eta_1^2\,.
\end{eqnarray}
The $\pm$ corresponds to the sign of $J+\bar{J}$ as it arises when the
square root is taken.  Coefficients denoted by Greek letters are
relative quantum variables defined in the appendix or earlier in this
section.  From this equation, it does not follow in an obvious manner
that there should be a bounce. As it stands, this equation may have no
solution for $\dot{a}$ --- or several ones.  Also here, we clearly see
the additional contribution by the squeezing term which is essential
to decide about the bounce, i.e.\ to find out where, if at all,
$\dot{a}=0$. If there would be no squeezing, i.e.\ $\eta_1=0=\eta_2$,
then the scale factor would bounce for an energy density near the
critical value of the free model, only slightly corrected by
$\epsilon_1$, $\delta_1$ and $\sigma$.

Instead of using the momentum and potential of $\phi$, we can express
the effective Friedmann equation in terms of energy density and
pressure $P$:
\begin{eqnarray}
 \left(\frac{\dot{a}}{a}\right)^2 &=& \frac{8\pi G}{3}
 \left(\rho+\frac{1}{2}\epsilon_1 (\rho-P)- \xi_2\frac{(\rho-P)^2}{\rho+P}
 \right)\nonumber\\
&&\qquad\times \left(1-\frac{1}{\rho_{\rm crit}}
 \left(\rho+\frac{\delta_1}{2}(\rho-P)+ 
\frac{\xi_1}{4}\frac{(\rho-P)^2}{\rho+P}
\right)+\sigma \right)
\nonumber\\
 && \pm\frac{2\pi G}{3}  (\rho-P)
\sqrt{1-\frac{1}{\rho_{\rm crit}}
 \left(\rho+\frac{\delta_1}{2}(\rho-P)+ 
\frac{\xi_1}{4}\frac{(\rho-P)^2}{\rho+P}
\right)+\sigma} \nonumber\\
&&\qquad\times
\left(\eta_1+\frac{\xi_3}{2} \frac{(\rho-P)^2}{\rho+P}\right)\nonumber\\
 && + \frac{2\pi G}{3}\frac{(\rho-P)^2}{\rho+P}
 \eta_1^2
\end{eqnarray}
where $\rho_{\rm crit}=3/8\pi G\gamma^2f_0^2a^{2+4x}$ is the same as
in the free case. However, $\rho=\rho_{\rm crit}$ would give a bounce
only in special cases.

\section{Conclusions}

We have derived effective equations for isotropic models in loop
quantum cosmology sourced by self-interacting or massive matter. In
contrast to models studied so far in this context, this introduces
quantum back-reaction which implies that the changing shape of a
quantum state influences the motion of its expectation values. This
effect, while initially negligible if one starts with a semiclassical
state, is important for the long-term evolution of cosmology. The
question of whether the classical singularity is replaced by a bounce
can be reliably addressed only if quantum back-reaction is
understood. Also the discussion of cyclic scenarios for the universe,
or any situation of long evolution times, requires these terms to be
taken into account.

We have derived perturbative effective equations which include quantum
back-reaction terms, based on the availability of the exactly solvable
model of \cite{BouncePert} as free system. Compared to an effective
Friedmann equation, which was previously available for models sourced
by a free, massless scalar, there are new terms when the scalar has a
potential.  Phenomenologically, this implies that the effective
Friedmann equation depends on the pressure of matter and not just on
its energy density as in the classical or free quantum case. Compared
to the free equation additional terms, which are independent of $f_0$,
do not arise from discreteness or repulsive gravity but from genuine
quantum back-reaction. These terms do grow in long-term evolution and
are especially important where the free system would bounce. We have
proven that there is still a bounce provided that the state at small
volume is uncorrelated, which presents a substantial generalization of
what has been available so far. However, we have also demonstrated
that the bounce is affected by quantum back-reaction of correlations,
and much more work is required to see how generically a bounce is
realized in this case. The now available equations, including those
for moments, are suitable for detailed numerical studies of the whole
parameter space. However, the parameters relevant for the bounce are
exactly the ones subject to cosmic forgetfulness
\cite{BeforeBB,Harmonic}, such that a firm and robust statement about
the bounce may be difficult to achieve.

Moreover, our quantum back-reaction equations show that a generic
state does develop strong quantum properties, especially correlations,
which are important for the space-time picture at small volume and
once a high-curvature region is traversed. This implies that
space-time does not stay as classical as detailed investigations of
the free model suggested; quantum geometry in general does become
important. It makes itself noticeable not just in repulsive forces but
also in genuine quantum variables such as fluctuations and
correlations. While quantum evolution remains non-singular at a basic
level of the whole wave function, in general there may not be a smooth
geometrical picture such as a simple bounce of only the classical
volume where genuine quantum variables would play no role.

For isotropic models, there are several places where a large matter
content $H$ helps to reduce the effects of quantum physics, thus
making the evolution appear more classical. (But note that the precise
rate of reduction depends on the refinement scheme of loop quantum
gravity which is only effectively included in isotropic models.) There
are, however, two important properties which prevent one from using
this for a general physical conclusion to the extent that effects of
quantum variables may be weak. First, already in isotropic models, a
large matter content enhances the sensitivity to quantum properties of
an initial state \cite{Harmonic} which, in long-term evolution, again
makes quantum effects play a large role. Secondly, the scale of
quantum effects is determined by the {\em total} matter content in the
universe only for homogeneous models, while inhomogeneous situations
are based on more local variables \cite{InhomLattice}. Those variables
would generically be much smaller than the total matter content unless
the spatial size of the universe is very tiny and comparable to the
Planck volume. In either case, one cannot escape strong quantum
effects when the small-volume behavior of cosmological models is
considered. Thus, the behavior for small $H$ is also of vital interest
even though an isotropic model for a large universe would require
large $H$. Similarly to a single massive partile in quantum
mechanics, which does not show strong quantum properties in contrast
to the microscopic dynamics of its elementary constituents, quantum
aspects of a large isotropic universe seem suppressed by its large
matter content. However, an isotropic model does not fully capture the
microscopic quantum gravitational dynamics, and it is this dynamics
which must be shown to be non-singular. Within isotropic models one
can obtain indications by looking at their properties also for small
$H$. While the basic removal of singularities based on an
extendability of wave functions has been shown also for inhomogeneous
systems of spherical symmetry \cite{SphSymmSing}, which thus captures
microscopic dynamics, this may in general not correspond to a smooth
bounce.

\bigskip

\noindent {\bf Acknowledgements:}
This work was supported in part by NSF grant PHY0653127.

\section*{Appendices}

\begin{appendix}

\section{Some useful Poisson brackets}
\label{a:Poisson}

For equations of motion of quantum variables, we are using Poisson
brackets
\begin{eqnarray*}
 \{C_{VJ},\langle\hat{V}\rangle\} &=& iC_{VJ} \quad,\quad
 \{C_{VJ},\langle\hat{J}\rangle\}= -i(\Delta J)^2\quad, \quad
 \{C_{VJ},\langle\hat{J}^{\dagger}\rangle\}= iC_{J\bar{J}}+ 2i(\Delta
 V)^2\\
 \{C_{VJ},(\Delta V)^2\} &=& 2iG^{VVJ} +2i\langle\hat{J}\rangle
 (\Delta V)^2- \frac{1}{6}i\hbar^2\langle\hat{J}\rangle\\
 \{C_{VJ},C_{V\bar{J}}\} &=& 4iG^{VVV} +6i(\langle\hat{V}\rangle+
{\textstyle\frac{1}{2}}\hbar) (\Delta V)^2-
i\langle\hat{J}\rangle C_{V\bar{J}}- i\langle\hat{J}^{\dagger}\rangle
C_{VJ}+ \frac{1}{2}i\hbar^2\langle\hat{V}\rangle+
\frac{1}{4}i\hbar^3\\
 \{C_{VJ},(\Delta J)^2\} &=& -2iG^{JJJ}- 2i\langle\hat{J}\rangle
 (\Delta J)^2\\
 \{C_{VJ},C_{J\bar{J}}\} &=& 2iG^{VVJ} +2i(\langle\hat{V}\rangle+
{\textstyle\frac{1}{2}}\hbar) C_{VJ}-
i\langle\hat{J}\rangle C_{J\bar{J}}+ i\langle\hat{J}^{\dagger}\rangle
(\Delta J)^2- \frac{1}{6}i\hbar^2\langle\hat{J}\rangle\\
 \{C_{VJ},(\Delta \bar{J})^2\} &=& 6iG^{VV\bar{J}}+ 8i(\langle\hat{V}\rangle+
{\textstyle\frac{1}{2}}\hbar) C_{V\bar{J}}- 2i\langle\hat{J}\rangle
(\Delta \bar{J})^2- i\hbar^2 \langle\hat{J}^{\dagger}\rangle
\end{eqnarray*}
for $C_{VJ}$ and
\begin{eqnarray*}
 \{(\Delta J)^2,\langle\hat{V}\rangle\} &=& 2i(\Delta J)^2 \quad,\quad
 \{(\Delta J)^2,\langle\hat{J}\rangle\} = 0 \quad,\quad \{(\Delta
 J)^2,\langle\hat{J}^{\dagger}\rangle\} = 4iC_{VJ}\\
 \{(\Delta J)^2,(\Delta V)^2\} &=& 4iG^{VJJ} +4i\langle\hat{J}\rangle
 C_{VJ}\\
 \{(\Delta J)^2,C_{VJ}\} &=& 2iG^{JJJ} +2i\langle\hat{J}\rangle
 (\Delta J)^2\\
\{(\Delta J)^2,C_{V\bar{J}}\} &=& 6iG^{VVJ}+
 8i(\langle\hat{V}\rangle+ 
{\textstyle\frac{1}{2}}\hbar) C_{VJ}-
2i\langle\hat{J}^{\dagger}\rangle (\Delta J)^2- i\hbar^2\langle\hat{J}\rangle\\
 \{(\Delta J)^2,C_{J\bar{J}}\} &=& 4iG^{VJJ}+ 4i (\langle\hat{V}\rangle+
{\textstyle\frac{1}{2}}\hbar) (\Delta J)^2\\
 \{(\Delta J)^2, (\Delta \bar{J})^2\} &=& 8iG^{VVV}+ 16i
 (\langle\hat{V}\rangle+ {\textstyle\frac{1}{2}}\hbar) (\Delta V)^2
 +8i(\langle\hat{V}\rangle+ 
{\textstyle\frac{1}{2}}\hbar) C_{J\bar{J}}- 8i\langle\hat{J}\rangle
C_{V\bar{J}}\\
 &&- 8i\langle\hat{J}^{\dagger}\rangle C_{VJ}+
4i\hbar^2\langle\hat{V}\rangle+ 2i\hbar^3
\end{eqnarray*}
for $(\Delta J)^2$. They can be derived from expectation values of
quantum commutators as discussed in more detail in \cite{BouncePot}.

\section{Expansion to second order in the potential}
\label{a:second}

To second order in the potential, we have the classical Hamiltonian
\begin{eqnarray*}
 H&=&\sqrt{-\frac{1}{4}(J-\bar{J})^2- V^{3/(1-x)}W(\phi)}\\
 &=& \frac{J-\bar{J}}{2i}- i\frac{V^{3/(1-x)}}{(J-\bar{J})^2} W+i
 \frac{V^{6/(1-x)}}{(J-\bar{J})^3}W^2+\cdots
\end{eqnarray*}
which implies the quantum Hamiltonian
\begin{eqnarray}
 H_Q &=& H_Q^{\rm lin}+ W^2\left(i\frac{V^{6/(1-x)}}{(J-\bar{J})^3}+
 3i\frac{5+x}{(1-x)^2} \frac{V^{2(2+x)/(1-x)}}{(J-\bar{J})^3} (\Delta
 V)^2\right.\\
 &&- \left.\frac{18i}{1-x} \frac{V^{(5+x)(1-x)}}{(J-\bar{J})^4}
 (C_{VJ}-C_{V\bar{J}})+ 6i\frac{V^{6/(1-x)}}{(J-\bar{J})^5} ((\Delta
 J)^2- 2C_{J\bar{J}}+ (\Delta \bar{J})^2)\right)\nonumber\\
 &=& \frac{J-\bar{J}}{2i}-i \frac{V^{3/(1-x)}}{J-\bar{J}}
 (1+\delta_1)W+ i\frac{V^{6/(1-x)}}{(J-\bar{J})^3} (1+\delta_2)W^2
\end{eqnarray}
where ``lin'' denotes terms linear in $W$ already presented in the
main text. In the last line, we have defined
\begin{equation}
 \delta_2 := 3\frac{5+x}{(1-x)^2}
\frac{(\Delta V)^2}{\langle\hat{V}\rangle^2}+
6\frac{(\Delta J)^2-2C_{J\bar{J}}+
(\Delta\bar{J})^2}{(\langle\hat{J}\rangle-
\langle\hat{J}^{\dagger}\rangle)^2}- \frac{18}{1-x}
\frac{C_{VJ}-C_{V\bar{J}}}{\langle\hat{V}\rangle
  (\langle\hat{J}\rangle- \langle\hat{J}^{\dagger}\rangle)}
\end{equation}
in addition to $\delta_1$ in (\ref{delta1}).

From this Hamiltonian, we derive the equation of motion
\begin{eqnarray}
 \frac{\md V}{\md\phi} &=& -\frac{J+\bar{J}}{2}+
 \frac{J+\bar{J}}{(J-\bar{J})^2} V^{3/(1-x)} (1+\epsilon_1) W- 3
 \frac{J+\bar{J}}{(J-\bar{J})^4} V^{6/(1-x)} (1+\epsilon_2)W^2\nonumber\\
&&+
 \frac{V^{(4-x)/(1-x)}}{(J-\bar{J})^2} \eta_1W-
 3\frac{V^{(7-x)/(1-x)}}{(J-\bar{J})^4}\eta_2W^2 \label{dVdphisecond}
\end{eqnarray}
with
\begin{eqnarray}
 \epsilon_2&:=& 3\frac{5+x}{(1-x)^2}
\frac{(\Delta V)^2}{\langle\hat{V}\rangle^2}+
10\frac{((\Delta J)^2-2C_{J\bar{J}}+
\Delta\bar{J})^2}{(\langle\hat{J}\rangle-
\langle\hat{J}^{\dagger}\rangle)^2}- \frac{24}{1-x}
\frac{C_{VJ}-C_{V\bar{J}}}{\langle\hat{V}\rangle
  (\langle\hat{J}\rangle- \langle\hat{J}^{\dagger}\rangle)}\,,\\
\eta_2 &:=& -4 \frac{(\Delta J)^2-
   (\Delta\bar{J})^2}{\langle\hat{V}\rangle(\langle\hat{J}\rangle-
\langle\hat{J}^{\dagger}\rangle)}+ \frac{6}{1-x}
\frac{C_{VJ}+C_{V\bar{J}}}{\langle\hat{V}\rangle^2}
\end{eqnarray}
in addition to $\epsilon_1$ in (\ref{eps}) and $\eta_1$ in (\ref{eta1}).

The equality $H_{\rm eff}=p_{\phi}$ now implies a fourth-order polynomial
for $J-\bar{J}$. To quadratic order in $W$, its perturbative solution is
\begin{equation}
 \frac{J-\bar{J}}{2i} = p_{\phi} \left(1+\frac{1}{2}
 \frac{V^{3/(1-x)}}{p_{\phi}^2} (1+\delta_1)W- \frac{1}{4}
 \frac{V^{6/(1-x)}}{p_{\phi}^4}
 (1+2\delta_1+\delta_1^2-\delta_2/2)W^2\right)\,.
\end{equation}
Inserting this into (\ref{dVdphisecond}) and using (\ref{trig}), we
obtain
\begin{eqnarray}
 \frac{1}{V^2}\left(\frac{\md V}{\md\phi}\right)^2 &=& \left(1-
 \frac{p_{\phi}^2}{V^2}- V^{(1+2x)/(1-x)} (1+\delta_1)W- \frac{1}{4}
 \frac{V^{2(2+x)/(1-x)}}{p_{\phi}^2} \xi_1 W^2+\sigma\right)\nonumber\\
 &&\qquad\times\left(1+\frac{V^{3/(1-x)}}{p_{\phi}^2} (1+\epsilon_1)W-
 \frac{V^{6/(1-x)}}{p_{\phi}^4} \xi_2 W^2\right)\nonumber\\
 &&\pm\frac{1}{2}\sqrt{1-
 \frac{p_{\phi}^2}{V^2}- V^{(1+2x)/(1-x)} (1+\delta_1)W- \frac{1}{4}
 \frac{V^{2(2+x)/(1-x)}}{p_{\phi}^2} \xi_1 W^2+\sigma}\nonumber\\
 &&\qquad\times \frac{V^{3/(1-x)} W}{p_{\phi}^2} \left(\eta_1
+\frac{1}{2}\xi_3 \frac{V^{3/(1-x)}W}{p_{\phi}^2}\right)\nonumber\\
 &&+ \frac{V^{6/(1-x)}}{4p_{\phi}^4}\eta_1^2W^2
\end{eqnarray}
with
\begin{eqnarray}
 \xi_1 &:=& 1+2\delta_1+\delta_2+\delta_1^2\,,\\
 \xi_2 &:=& \frac{1}{2}\epsilon_1+\delta_1-\frac{3}{4}\epsilon_2+
 \epsilon_1\delta_2-\frac{1}{4}\epsilon_1^2\,,\\
 \xi_3 &:=& -\eta_1-\frac{3}{2}\eta_2+\eta_1\epsilon_1-2\eta_1\delta_1\,.
\end{eqnarray}
Reinstating numerical factors, this provides the effective Friedmann
equation (\ref{EffFriedQBR}).

\end{appendix}


\end{document}